\def\year{2022}\relax
\documentclass[letterpaper]{article} % DO NOT CHANGE THIS
\usepackage{aaai22}  % DO NOT CHANGE THIS
\usepackage{times}  % DO NOT CHANGE THIS
\usepackage{helvet}  % DO NOT CHANGE THIS
\usepackage{courier}  % DO NOT CHANGE THIS
\usepackage[hyphens]{url}  % DO NOT CHANGE THIS
\usepackage{graphicx} % DO NOT CHANGE THIS
\usepackage{natbib}  % DO NOT CHANGE THIS AND DO NOT ADD ANY OPTIONS TO IT
\usepackage{caption} % DO NOT CHANGE THIS AND DO NOT ADD ANY OPTIONS TO IT
\DeclareCaptionStyle{ruled}{labelfont=normalfont,labelsep=colon,strut=off} % DO NOT CHANGE THIS
\usepackage{algorithm}
\usepackage{algorithmic}

\usepackage{booktabs} 
\usepackage{amsmath}
\usepackage{amssymb}
\usepackage{hyperref} 
\usepackage{multirow}

\usepackage{newfloat}
\usepackage{listings}
\floatstyle{ruled}
\newfloat{listing}{tb}{lst}{}
\floatname{listing}{Listing}
\title{A Recommender System for NFT Collectibles with Item Feature}
\author{
    \\
    Minjoo Choi\textsuperscript{\rm 1}\equalcontrib,
    Seonmi Kim\textsuperscript{\rm 1}\equalcontrib,
    Yejin Kim\textsuperscript{\rm 1}\equalcontrib,
    Youngbin Lee\textsuperscript{\rm 1}\equalcontrib,
    Joohwan
    Hong\textsuperscript{\rm1}\thanks{Corresponding author},
    Yongjae Lee\textsuperscript{\rm 1}\thanks{Corresponding author}
}
\affiliations{
    \textsuperscript{\rm 1}Ulsan National Institute of Science and Technology(UNIST)
    \{mincule, msraask3, kimyejin99, young, joohwanhong, yongjaelee\}@unist.ac.kr
}

\begin{document}

\maketitle

\begin{abstract}
Recommender systems have been actively studied and applied in various domains to deal with information overload. Although there are numerous studies on recommender systems for movies, music, and e-commerce, comparatively less attention has been paid to the recommender system for NFTs despite the continuous growth of the NFT market. This paper presents a recommender system for NFTs that utilizes a variety of data sources, from NFT transaction records to external item features, to generate precise recommendations that cater to individual preferences. We develop a data-efficient graph-based recommender system to efficiently capture the complex relationship between each item and users and generate node(item) embeddings which incorporate both node feature information and graph structure. Furthermore, we exploit inputs beyond user-item interactions, such as image feature, text feature, and price feature. Numerical experiments verify the performance of the graph-based recommender system improves significantly after utilizing all types of item features as side information, thereby outperforming all other baselines.
\end{abstract}

\section{1. Introduction}

Recommender systems are a subset of artificial intelligence used to browse a large collection of objects and make suggestions for products, services, or content. The past few years have witnessed major advances in recommender systems as the interest in harnessing deep learning methods for recommender systems surged. The representations generated from deep learning models are proven to be useful as they capture complex nonlinear relationships between users and items, thereby outperforming traditional methods including Matrix Factorization and Collaborative Filtering. Among deep learning models, one area is graph-based models, which efficiently deal with graph-structured data like user-item interactions or relationships between entities. The most remarkable advantages of these models are their capability of exploring high-order connectivity and leveraging the information of the local graph neighborhoods \cite{he2020lightgcn, wang2019neural}. They have been actively studied in various domains and have outperformed countless recommender systems, however, graph-based recommender systems for Non-Fungible Tokens (NFT) are yet to be explored. 

Where cryptocurrency was proposed as the digital substitute for currency, NFTs are nowadays being touted as the digital substitute for collectibles. With NFTs being a growing trend, the total worth of NFT market reached \$41 million in 2021 alone, nearing that of the entire global fine art market \cite{Conti2022nftexp}. It was quite recently, at the beginning of 2020, when the NFT market started to take off \cite{vairagade2022proposal}. There were 1,415,638 NFT sales with a trading volume of 82,492,916 USD in 2020, which later soared to 27,414,477 sales and a trading volume of 17,694,851,721 USD in 2021 \cite{besancia2022nftreport}. Among them, most of the transactions identified are for collectibles which take up to 27\% of the total sales. As it is a market with great potential, and every transaction information is digitized and disclosed, we expect that there will be considerable benefits from applying the recommendation system.

 However, there are a few challenges specific to NFT recommender systems. NFTs are absolutely lacking in feedback information since NFT’s uniqueness allows only one person to own it at a time while music, movies, and products on online platforms can be consumed by multiple users simultaneously and therefore are more likely to have plenty amount of feedback information. In addition, it seems to be very difficult to specify user features since NFTs operate in a blockchain system, which is based on anonymity. 
 
 In this paper, we implement a graph-based recommender system which also incorporates item features (NFT images, textual information, and price) so as to tackle the aforementioned challenges. Graph learning provides a unified way to utilize all available different types of data and effectively capture non-trivial user-item, user-user, and item-item relations. Our experimental results confirm these expectations, graph-based models significantly improving the recommendation performance by effectively incorporating all types of item features and outperforming other models.

\section{2. Related Works}
Non-Fungible Token (NFT) is a new type of cryptocurrency that represents a unique digital asset such as metaverse, collectibles, music, and even game items. In this paper we focus on NFT collectibles.

The majority of NFT collections feature a large number of NFT items that all share a similar artistic style. A decent example of this would be Bored Apes Yacht Club (BAYC), a collection made up of 10,000 NFT artworks depicting variants of Bored Apes. Every token in BAYC shares same array of traits including ‘fur’, and ‘clothes’, depicting the color and style of its fur, and outfit of the Bored Ape, respectively. The rarer each token’s features, the higher value it is prone to fetch. That said, NFT collections can take on many different forms, such as commemorative tickets, music albums, and so on.

\subsubsection{Recommender Systems}

Broadly speaking, there are two main approaches in recommender systems. The content-based methods rely on the characteristics of the items historically consumed by a user whereas the Collaborative filtering (CF) requires users’ interactions and feedback on a set of items.

Recommendation system expands its area to deep learning to capture higher-order interactions in the data. Especially, recent years have seen major advancements in graph-based recommender system, which is efficient in modelling complex relationship between users and items in a graph and provide recommendations based on graph exploration \cite{zheng2018spectral, wang2019neural}.
% \cite{bruna2013spectral, zheng2018spectral, wang2019neural}
Recently proposed method Graph Convolutional Network \cite{kipf2016semi} has developed a graph convolution, followed by numerous works that adapt GCN to recommendation system \cite{berg2017graph, zheng2018spectral, wang2019neural, he2020lightgcn, hamilton2017inductive, ying2018graph}. The underlying structure for most of the GCN based recommendation system is user-item collaborative filtering \cite{zheng2018spectral,wang2019neural}. SpectralCF \cite{zheng2018spectral} operates a spectral convolution in the spectral domain in order to find the proximity and all possible connectivity between users and items. NGCF \cite{wang2019neural} proposes iterative user and item embeddings propagation to add collaborative signal into the embedding process.

\subsubsection{NFT Recommender Systems}

The idea of applying recommender system to NFT sector was first proposed in a blog article, where they presented a multiple regression based recommender system to help users to navigate recommendations voluntarily \cite{opensea2020}. \citeauthor{pradana2022multi} \shortcite{pradana2022multi} proposed a multi-criteria recommender system for NFT-based IAP in RPG game while \citeauthor{piyadigama2022analysis} \shortcite{piyadigama2022analysis} identified some meaningful features for NFT recommender systems and proposed a method to integrate social trends into NFT recommendations \cite{piyadigama2022exploration}. However, these studies left room for further improvement as they did not consider the unique characteristics of NFTs when building a recommender system for NFTs. In addition, to our knowledge, this is the first study to apply a graph-based recommender system to exploit high-order connectivity between users and items effectively.

\section{3. Data} \label{Data}

\subsubsection{Data Preparation}

We selected five collections, BAYC, Azuki, Meebits, Doodles, and Cool Cats, which were ranked within the top 15 in terms of market capitalization as of the end of August 2022 on \citeauthor{dappradar2022}.

We collected about a year-long transaction data from September 2021 to September 2022 from Etherscan NFT tracker \cite{etherscan2022} and the transaction data consists of transaction hash, buyer's wallet address, token ID, and price of the token sold.

\subsubsection{Data Description and Sparsity}
Our data consists of implicit feedback that only considers purchase history as interactions. Items in our user-item matrices have at least 3 interactions. Typically benchmark data set a higher threshold of 5 or 10, however, considering the lack of transaction history due to the uniqueness of NFTs and the fact that the NFT market is still in its infancy, we have set relatively low threshold. 
Specific figures for the user-item matrix of each collection can be found in the Table \ref{data_description}.

In addition to that, we also included some additional features of NFTs to address the transaction data shortage and to boost the predictive power of the recommender system. The description of each type of feature can be found in subsequent paragraphs of this section.

\begin{table}[t]
\centering

\begin{tabular}{l|c|c|c}
\toprule
Collection & Users & Items & Interactions \\
% \hline
% \hline

\midrule
BAYC & 2,469 & 929 & 3,504 \\
Azuki & 9,384 & 3,931 & 16,401 \\
Meebits & 4,581 & 1,694 & 7,317 \\
Doodles & 9,321 & 3,497 & 15,176 \\
Cool Cats & 4,806 & 1,647 & 6,667 \\
\end{tabular}

\caption{NFT data description}
\label{data_description}
\end{table}

\subsection{Item Features}
Besides transaction data shortage, as alluded to earlier, the anonymity of the blockchain makes it difficult to specify the user's characteristics. We therefore collected external features and price of each NFT to compensate for the shortcomings specific to NFT data.

\subsubsection{Image}

We downloaded images from the url of each token we found using OpenSea API \cite{openseaapi2022}. Each collection has about 10,000 images in general.

\subsubsection{Text}

We have sourced text data from OpenSea API. Text data is comprised of tags describing visual properties like ‘background color’, ‘body’, and ‘outfit’ of each NFT (token).

\subsubsection{Price}

Each item's average selling price was used as a price feature of the item. We only considered the selling price stated in Ethereum (ETH) and Wrapped Ethereum (WETH) and replaced the rest with 0.

\section{4. Method}

\subsubsection{Problem Definition}

Each wallet address represents each user $u$, and each token ID as each item $i$, where set of users is $U = \{u_{1},u_{2},\dots,u_{N}\}$, and set of items is $I = \{i_{1},i_{2},\dots,i_{M}\}$. A user-item interaction matrix $\mathbf{Y} \in \mathbb{R}^{N \times M}$ consists of implicit feedback, where $y_{ui}=1$ if a user bought an item, and $y_{ui}=0$, otherwise. The matrix $\mathbf{Y}$ is then converted into a bipartite graph.

Since our goal is to predict whether the user has a potential interest in item $v$, we aim to learn a prediction function $\hat {y}_{ui}= \mathcal F(u,i|\Theta,\mathbf{X},\mathbf{Y}) $ where $\hat {y}_{ui}$ denotes the probability that user $u$ will buy item $i$, $\mathbf{X}$ denotes combined representations of three different features $\mathbf{X}_{img}, \mathbf{X}_{txt}, \mathbf{X}_{price}$ that are described in \nameref{Data} section, and $\Theta$ denotes the trainable model parameters of function $\mathcal F$.

\subsubsection{Item Feature Embedding}

As a first step, we make a representation of each of the raw item feature data, which will be the input for the model. Image representations are obtained using convolutional auto-encoders (CAEs) \cite{masci2011stacked}. An encoder $f$ maps our input $\mathbf{X}_{img}$ to the latent representation $\mathbf{X^*}_{img}$, and a decoder $g$ generates reconstructed input $\mathbf{X'}_{img}$. Once the model is trained to minimize MSE, only the encoder part is used to extract representations. That is, by using 3-layer encoder $f$ with 3*3 convolutional kernels, 2*2 max pooling and flattening, we get $\mathbf{X^*}_{img}\in\mathbb{R}^{M\times 64}$.

We converted each word into 300-dimensional word embedding by retrieving the embedding of each word present in the pre-trained Word2Vec model \cite{mikolov2013distributed} and otherwise replaced it with a 300-dimensional vector of zeros.
\footnote{Most of the data describing each visual property consists of a single word, however, in case there are multiple words describing a single property like ‘short red hair’ for ‘hair’, a sum of each word embedding was used instead.}
Each word embedding was then concatenated with other embeddings corresponding to the same item, with each item's word embedding size varying between 1500 and 1800 depending on the number of visual traits considered. As a result, we obtain textual information representation, $\mathbf{X^*}_{txt}\in\mathbb{R}^{M\times 1500or1800}$

The size of the price feature vector is initially $\mathbf{X}_{price}\in\mathbb{R}^{M\times 1}$. However, because the dimension of the item representation becomes the dimension of the item node embedding in the graph, we need the size of the price representation to be large enough to contain some information. Therefore, we replicate the original price feature vector and then concatenate those to get a larger representation of $\mathbf{X^*}_{price}\in\mathbb{R}^{M\times 64}$.

\subsection{Model Architecture}

Our framework uses three different types of item features into the \textbf{Neural Graph Collaborative Filtering (NGCF)}. Thus, we follow the model architecture and the notations of NGCF \cite{wang2019neural}.

\subsubsection{ (1) Input Layer}

For NGCF-all model, all types of NFT features for each item are concatenated: $\mathbf{X^*}
=
\mathbf{X^*}_{img}
\| 
\mathbf{X^*}_{txt}
\| 
\mathbf{X^*}_{price}$ and projected to a low-dimensional latent space using three linear layers of Multi-Layer Perceptrons(MLPs): $\mathbf{X}
=
MLP(\mathbf{X^*})$. On the other hand, we also made a model utilizing one type of feature each: NGCF-image, NGCF-text, and NGCF-price.

\subsubsection{(2) Embedding Layer}

$N$ users and $M$ items are embedded into $64$-dimensional vectors, $\mathbf{e}_{u_1}, \cdots, \mathbf{e}_{u_N}$, $\mathbf{e}_{i_1}, \cdots, \mathbf{e}_{i_M}$, respectively. User embeddings are then initialized with Xavier initialization method whereas item embeddings are initialized with item representations we extracted earlier. In other words, the weight matrix of the item embedding layer is initialized with the item representation matrix $\mathbf{X}$.

\subsubsection{(3) Embedding Propagation Layers}

User and item embeddings are refined by GCN-based propagation. For example, a node embedding at $l$-th layer receives messages $\mathbf{m}$ from their neighbors and itself: $\mathbf{e}_u^{(l)}=\operatorname{LeakyReLU}\left( \mathbf{m}_{u}^{(l)} \right)$, where $\mathbf{m}$ is defined in the same way as \citet{wang2019neural}.

\subsubsection{(4) Prediction Layer}

Refined embeddings from each propagation layers are then aggregated to create the final embeddings $\mathbf{e}^* = \mathbf{e}^{(0)}\|\cdots\| \mathbf{e}^{(L)}$ and the predicted value of user’s preference is calculated by taking the inner product of the final embeddings: $\hat{y}_{ui}=\mathbf{e}_u^{* \top} \mathbf{e}_i^*$. For optimization, a pairwise BPR loss \cite{rendle2012bpr} is used to learn parameters in (1) Input Layer, (2) Embedding Layer, and (3) Embedding Propagation Layers.

\section{5. Experiment}

\begin{table*}[t]
\centering
\resizebox{\textwidth}{!}{%
\begin{tabular}{|l|c|c|c|c|c|c|c|c|c|c|c|c|}
\hline
\multirow{2}{*}{Methods} &
\multicolumn{2}{c|}{Azuki} &
\multicolumn{2}{c|}{BAYC} &
\multicolumn{2}{c|}{Cool Cats} &
\multicolumn{2}{c|}{Doodles} &
\multicolumn{2}{c|}{Meebits}&
\multicolumn{2}{c|}{Average}\\
\cline{2-13}
& {Recall} & {NDCG} &
{Recall} & {NDCG} &
{Recall} & {NDCG} &
{Recall} & {NDCG} &
{Recall} & {NDCG} &
{Recall} & {NDCG} \\
\hline

Pop & 0.0718 & 0.0318 & 0.1152 & 0.0518 & 0.0741 & 0.0360 & 0.0755 & 0.0351 & 0.0692 & 0.0284 & 0.0812 & 0.0366\\
ItemKNN & 0.1027 & 0.0486 & 0.1217 & 0.0592 & 0.1016 & 0.0475 & 0.0978 & 0.0485 & 0.1192 & 0.0654 & 0.1086 & 0.0538\\
BPR & 0.0492 & 0.0443 & 0.0971 & 0.0358 & 0.1238 & 0.0532 & 0.0937 & 0.0432 & 0.1134 & 0.0562 & 0.1058 & 0.0465\\
DMF & 0.0975 & 0.0463 & 0.1111 & 0.0466 & 0.1196 & 0.0540 & 0.1046 & 0.0493 & 0.1047 & 0.0492 & 0.1075 & 0.0491\\
NeuMF & 0.0849 & 0.0385 & 0.0864 & 0.0367 & 0.0905 & 0.0412 & 0.0970 & 0.0429 & 0.0841 & 0.0455 & 0.0886 & 0.0410\\
NGCF & 0.1046 & 0.0486 & 0.1079 & 0.0480 & 0.1040 & 0.0480 & 0.1017 & 0.0471 & 0.1443 & 0.0749 & 0.1125 & 0.0533\\
NGCF-img & 0.1200 & \textbf{0.0583} & 0.0975 & 0.0469 & 0.1113 & 0.0514 & 0.0914 & 0.0426 & 0.1501 & 0.0687 & 0.1141 & 0.0536\\
NGCF-price & 0.1220 & 0.0558 & 0.1089 & 0.0499 & 0.1545 & 0.0671 & \textbf{0.1616} & \textbf{0.0761} & 0.1425 & 0.0662 & 0.1379 & 0.0630\\
NGCF-txt & \textbf{0.1237} & 0.0567 & \textbf{0.1866} & \textbf{0.0895} & \textbf{0.1745} & \textbf{0.0775} & 0.1275 & 0.0603 & \textbf{0.1726} & \textbf{0.0803} & \textbf{0.1570} & \textbf{0.0729}\\
NGCF-all & \textbf{0.1548} & \textbf{0.0729} & \textbf{0.1474} & \textbf{0.0656} & \textbf{0.2287} & \textbf{0.1083} & \textbf{0.1554} & \textbf{0.0724} & \textbf{0.2193} & \textbf{0.1118} & \textbf{0.1811} & \textbf{0.0862}\\ 
% NGCF-sum64 & 0.1302 & 0.0605 & 0.1147 & 0.0493 & 0.1717 & 0.0764 & 0.1357 & 0.0649 & 0.1701 & 0.0787 & 0.1445 & \textbf{0.0791}\\ 
\hline
\end{tabular}
}
\caption{Results measured by Recall@10 and NDCG@10}
\label{results_baseline_ngcf}
\end{table*}

\begin{figure*}[t]
\hspace{1.2em} % Adjust the space as needed
\centerline{\includegraphics[width=2.65\columnwidth]{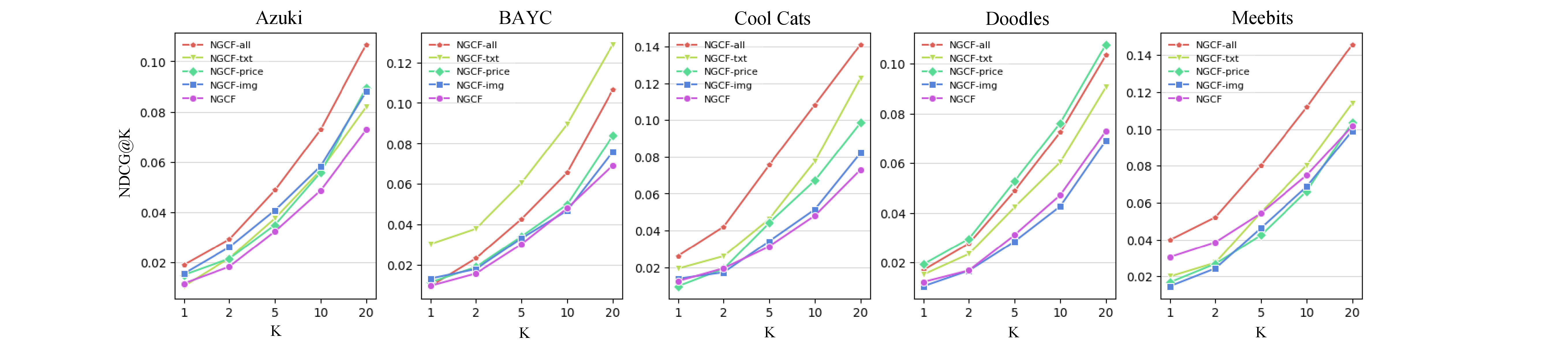}}
\caption{Performance comparison between NGCF and NGCF variant models over the top K recommendations on different datasets.}
\label{ndcg_plot}
\end{figure*}

We performed experiments on the NFT transaction dataset we collected. All the experiments are implemented based on an open-sourced recommendation
library, Recbole.

\subsubsection{Data sets}

We filtered transaction data and made five user-item matrices as discussed in \nameref{Data} section. Moreover, we conducted a negative sampling to make a user-specific pairwise preference. Items purchased by the user are set as positive.

\subsubsection{Evaluation schema}

 We used NDCG@K (Normalized Discounted Cumulative Gain) and Recall@K to evaluate the performance of the top $K$ recommendations. We randomly split the historical user-item interaction of each collection into training, validation and test sets with proportions of $80\%$, $10\%$, and $10\%$, respectively. 

\subsubsection{Model comparison}

We compared a basic NGCF with non graph-based models including Pop\cite{zhao2021recbole}, ItemKNN\cite{deshpande2004item}, BPR\cite{rendle2012bpr}, DMF\cite{xue2017deep}, NeuMF\cite{he2017neural}. We also conducted ablation studies and considered several NGCF variants utilizing different types of NFT features; NGCF-img, NGCF-txt, NGCF-price, and NGCF-all when evaluating the significance of each feature.

\subsubsection{Hyperparameter settings}

We optimized every model with Adam optimizer, where the batch size and epochs were fixed at $2048$ and $50$. In terms of hyperparameters, we applied a grid search and used NDCG@20 as an indicator. The search ranges of NGCF models are as follows: learning rate of $[0.01, 0.001]$, node dropout ratio in $[0.0, 0.2]$, message dropout ratio in $[0.0, 0.2]$, regularization weight of $[1e-5, 1e-3]$, and the number of Embedding Propagation Layers as $[1, 2, 3]$, where the embedding size is fixed at 512. We did hyperparameter tuning for the baseline models as well.

\subsection{Results and Discussion}
Table \ref{results_baseline_ngcf} shows the results of the NGCF, NGCF variants, and baseline models. %The two largest values are in bold.

\subsubsection{Effect of item features}

We start our analysis by comparing the performance of NGCF and its variant models. Figure \ref{ndcg_plot}, shows that both NGCF-txt and NGCF-all outperform NGCF without features across all collections while NGCF-img outperforms NGCF on 3 collections by a slight margin. NGCF-price outperforms NGCF on the majority of the collections and achieved the best performance on Doodles dataset. This demonstrates that tag information (text), and price are the most influential features of NFTs that shape user's preference whereas the image feature is less likely to be influential. In addition, from the fact that price feature enhances the recommendation performance, it seems like adding the price information of each product could reflect user's budgets.
%we can conclude that the users tend to purchase NFTs within the price range of the NFTs they purchased in the past. %than we thought aside from Azuki collection if we see the results of NGCF-img. 
%We can observe that NGCF-price is doing well in Azuki, Coolcats and Doodles collections.

%The results of applying features show that text and price information of an asset provides more suitable recommendation for users. It is presumed that tags of the asset explains users’ purchasing tendency. Price of the asset is also important in recommendation with investment purpose. In fact, it is quite difficult for individual investors to suddenly increase their purchase level within a short period of one year.

\subsubsection{Peformance comparison with the baselines}

Table \ref{results_baseline_ngcf} compares the performance of the various approaches.  The average recall score for all datasets shows NGCF and its variants consistently outperforming the baselines by a significant margin. This verifies the effectiveness of graph-based models, empirically demonstrating modelling high-order relations can well facilitate the recommendation task.
%The result indicates that NGCF and its variants are superior to others in all collections from which one can conclude that graph based recommender system with additional item features efficiently models the complex relationship between users and items and utilizes the neighborhood information.  

\section{6. Conclusion}
We proposed an NFT recommender system that utilizes item features to overcome the limitations stemming from the unique characteristics of NFTs. We implemented the recommender system on the actual NFT transactions data and evaluated the recommendation performance with offline metrics demonstrating substantial improvement in recommendation performance. Our work identified some useful NFT features for recommender systems and shows graph-based models efficiently leverage neighbor information.

\bibliography{aaai22.bib}

\begin{thebibliography}{24}
\providecommand{\natexlab}[1]{#1}

\bibitem[{Berg, Kipf, and Welling(2018)}]{berg2017graph}
Berg, R. v.~d.; Kipf, T.~N.; and Welling, M. 2018.
\newblock Graph convolutional matrix completion.
\newblock In \emph{Proceedings of the 24th ACM SIGKDD international conference on knowledge discovery \& data mining}.

\bibitem[{{Besancia}(2022)}]{besancia2022nftreport}
{Besancia}. 2022.
\newblock Our 2021 NFT Yearly Report is out!
\newblock \url{https://nonfungible.com/news/corporate/yearly-nft-market-report-2021}.
\newblock Accessed: 2022-10-25.

\bibitem[{{Conti and Schmidt}(2022)}]{Conti2022nftexp}
{Conti and Schmidt}. 2022.
\newblock What Is An NFT? Non-Fungible Tokens Explained.
\newblock \url{https://www.forbes.com/advisor/investing/cryptocurrency/nft-non-fungible-token/}.
\newblock Accessed: 2022-10-24.

\bibitem[{{DappRadar}(2022)}]{dappradar2022}
{DappRadar}. 2022.
\newblock DappRadar.
\newblock \url{https://dappradar.com/nft/collections/protocol/ethereum/1}.
\newblock Accessed: 2022-08-31.

\bibitem[{Deshpande and Karypis(2004)}]{deshpande2004item}
Deshpande, M.; and Karypis, G. 2004.
\newblock Item-based top-n recommendation algorithms.
\newblock \emph{ACM Transactions on Information Systems (TOIS)}, 22(1): 143--177.

\bibitem[{{Etherscan}(2022)}]{etherscan2022}
{Etherscan}. 2022.
\newblock NFT Tracker.
\newblock \url{https://etherscan.io/nfttracker}.
\newblock Accessed: 2022-09-30.

\bibitem[{Hamilton, Ying, and Leskovec(2017)}]{hamilton2017inductive}
Hamilton, W.; Ying, Z.; and Leskovec, J. 2017.
\newblock Inductive representation learning on large graphs.
\newblock \emph{Advances in neural information processing systems}, 30.

\bibitem[{He et~al.(2020)He, Deng, Wang, Li, Zhang, and Wang}]{he2020lightgcn}
He, X.; Deng, K.; Wang, X.; Li, Y.; Zhang, Y.; and Wang, M. 2020.
\newblock Lightgcn: Simplifying and powering graph convolution network for recommendation.
\newblock In \emph{Proceedings of the 43rd International ACM SIGIR conference on research and development in Information Retrieval}, 639--648.

\bibitem[{He et~al.(2017)He, Liao, Zhang, Nie, Hu, and Chua}]{he2017neural}
He, X.; Liao, L.; Zhang, H.; Nie, L.; Hu, X.; and Chua, T.-S. 2017.
\newblock Neural collaborative filtering.
\newblock In \emph{Proceedings of the 26th international conference on world wide web}, 173--182.

\bibitem[{Kipf and Welling(2017)}]{kipf2016semi}
Kipf, T.~N.; and Welling, M. 2017.
\newblock Semi-supervised classification with graph convolutional networks.
\newblock In \emph{International Conference on Learning Representations (ICLR) 2017}.

\bibitem[{Masci et~al.(2011)Masci, Meier, Cire{\c{s}}an, and Schmidhuber}]{masci2011stacked}
Masci, J.; Meier, U.; Cire{\c{s}}an, D.; and Schmidhuber, J. 2011.
\newblock Stacked convolutional auto-encoders for hierarchical feature extraction.
\newblock In \emph{International conference on artificial neural networks}, 52--59. Springer.

\bibitem[{Mikolov et~al.(2013)Mikolov, Sutskever, Chen, Corrado, and Dean}]{mikolov2013distributed}
Mikolov, T.; Sutskever, I.; Chen, K.; Corrado, G.~S.; and Dean, J. 2013.
\newblock Distributed representations of words and phrases and their compositionality.
\newblock \emph{Advances in neural information processing systems}, 26.

\bibitem[{{OpenSea}(2020)}]{opensea2020}
{OpenSea}. 2020.
\newblock What are you missing? Using basic machine learning to predict and recommend NFTs with OpenSea data.
\newblock \url{https://opensea.io/blog/guides/predict-and-recommend-nfts/}.
\newblock Accessed: 2022-10-24.

\bibitem[{{OpenSea}(2022)}]{openseaapi2022}
{OpenSea}. 2022.
\newblock OpenSea API.
\newblock \url{https://docs.opensea.io/reference/api-overview}.
\newblock Accessed: 2022-05-28.

\bibitem[{Piyadigama and Poravi(2022{\natexlab{a}})}]{piyadigama2022analysis}
Piyadigama, D.; and Poravi, G. 2022{\natexlab{a}}.
\newblock An Analysis of the Features Considerable for NFT Recommendations.
\newblock \emph{arXiv preprint arXiv:2205.00456}.

\bibitem[{Piyadigama and Poravi(2022{\natexlab{b}})}]{piyadigama2022exploration}
Piyadigama, D.~R.; and Poravi, G. 2022{\natexlab{b}}.
\newblock Exploration of the possibility of infusing Social Media Trends into generating NFT Recommendations.
\newblock \emph{arXiv preprint arXiv:2205.11229}.

\bibitem[{Pradana et~al.(2022)Pradana, Hariadi, Rachmadi, and Arif}]{pradana2022multi}
Pradana, R.~P.; Hariadi, M.; Rachmadi, R.~F.; and Arif, Y.~M. 2022.
\newblock A Multi-Criteria Recommender System For NFT Based IAP In RPG Game.
\newblock In \emph{2022 International Seminar on Intelligent Technology and Its Applications (ISITIA)}, 214--219. IEEE.

\bibitem[{Rendle et~al.(2009)Rendle, Freudenthaler, Gantner, and Schmidt-Thieme}]{rendle2012bpr}
Rendle, S.; Freudenthaler, C.; Gantner, Z.; and Schmidt-Thieme, L. 2009.
\newblock BPR: Bayesian personalized ranking from implicit feedback.
\newblock In \emph{The Conference on Uncertainty in Artificial Intelligence (UAI)}, 452--461.

\bibitem[{Vairagade et~al.(2022)Vairagade, Bitla, Judge, Dharpude, and Kekatpure}]{vairagade2022proposal}
Vairagade, R.; Bitla, L.; Judge, H.~H.; Dharpude, S.~D.; and Kekatpure, S.~S. 2022.
\newblock Proposal on NFT Minter for Blockchain-based Art-Work Trading System.
\newblock In \emph{2022 IEEE 11th International Conference on Communication Systems and Network Technologies (CSNT)}, 571--576. IEEE.

\bibitem[{Wang et~al.(2019)Wang, He, Wang, Feng, and Chua}]{wang2019neural}
Wang, X.; He, X.; Wang, M.; Feng, F.; and Chua, T.-S. 2019.
\newblock Neural graph collaborative filtering.
\newblock In \emph{Proceedings of the 42nd international ACM SIGIR conference on Research and development in Information Retrieval}, 165--174.

\bibitem[{Xue et~al.(2017)Xue, Dai, Zhang, Huang, and Chen}]{xue2017deep}
Xue, H.-J.; Dai, X.; Zhang, J.; Huang, S.; and Chen, J. 2017.
\newblock Deep matrix factorization models for recommender systems.
\newblock In \emph{IJCAI}, volume~17, 3203--3209. Melbourne, Australia.

\bibitem[{Ying et~al.(2018)Ying, He, Chen, Eksombatchai, Hamilton, and Leskovec}]{ying2018graph}
Ying, R.; He, R.; Chen, K.; Eksombatchai, P.; Hamilton, W.~L.; and Leskovec, J. 2018.
\newblock Graph convolutional neural networks for web-scale recommender systems.
\newblock In \emph{Proceedings of the 24th ACM SIGKDD international conference on knowledge discovery \& data mining}, 974--983.

\bibitem[{Zhao et~al.(2021)Zhao, Mu, Hou, Lin, Chen, Pan, Li, Lu, Wang, Tian et~al.}]{zhao2021recbole}
Zhao, W.~X.; Mu, S.; Hou, Y.; Lin, Z.; Chen, Y.; Pan, X.; Li, K.; Lu, Y.; Wang, H.; Tian, C.; et~al. 2021.
\newblock Recbole: Towards a unified, comprehensive and efficient framework for recommendation algorithms.
\newblock In \emph{Proceedings of the 30th ACM International Conference on Information \& Knowledge Management}, 4653--4664.

\bibitem[{Zheng et~al.(2018)Zheng, Lu, Jiang, Zhang, and Yu}]{zheng2018spectral}
Zheng, L.; Lu, C.-T.; Jiang, F.; Zhang, J.; and Yu, P.~S. 2018.
\newblock Spectral collaborative filtering.
\newblock In \emph{Proceedings of the 12th ACM conference on recommender systems}, 311--319.

\end{thebibliography}

\end{document}